# Scaled charges for ions: An improvement but not the final word for modeling electrolytes in water

Samuel Blazquez: Conceptualization equal, Data curation lead, Formal analysis equal, Investigation equal, Writing ‐ original draft equal, Writing ‐ review & editing equal, Maria M. Conde: Funding acquisition equal, Resources equal, Supervision equal, Writing ‐ review & editing equal, Carlos Vega: Conceptualization lead, Funding acquisition equal, Investigation equal, Methodology equal, Supervision equal, Writing ‐ original draft equal, Writing ‐ review & editing equal

S. Blazquez,[1] ORCID, M. M. Conde,[2] ORCID, and C. Vega[1,a)] ORCID

[1]Dpto. Química Física I, Fac. Ciencias Químicas, Universidad Complutense de Madrid, 28040 Madrid, Spain

[2]Departamento de Ingeniería Química Industrial y Medio Ambiente, Escuela Técnica Superior de Ingenieros Industriales, Universidad Politécnica de Madrid, 28006 Madrid, Spain

[a)] Author to whom correspondence should be addressed: cvega@quim.ucm.es

In this work, we discuss the use of scaled charges when developing force fields for NaCl in water. We shall develop force fields for Na$^+$ and Cl$^-$ using the following values for the scaled charge (in electron units): $\pm 0.75$, $\pm 0.80$, $\pm 0.85$, and $\pm 0.92$ along with the TIP4P/2005 model of water (for which previous force fields were proposed for q = $\pm 0.85$ and q = $\pm 1$). The properties considered in this work are: densities, structural properties, transport properties, surface tension, freezing point depression, and maximum in density. All the developed models were able to describe quite well the experimental values of the densities. Structural properties were well described by models with charges equal to or larger than ±0.85, surface tension by the charge ±0.92, maximum in density by the charge ±0.85, and transport properties by the charge ±0.75. The use of a scaled charge of ±0.75 is able to reproduce with high accuracy the viscosities and diffusion coefficients of NaCl solutions for the first time. We have also considered the case of KCl in water, and the results obtained were fully consistent with those of NaCl. There is no value of the scaled charge able to reproduce all the properties considered in this work. Although certainly scaled charges are not the final word in the development of force fields for electrolytes in water, its use may have some practical advantages. Certain values of the scaled charge could be the best option when the interest is to describe certain experimental properties.

# I. INTRODUCTION

Aqueous electrolyte solutions are of interest from both practical and theoretical points of view. Ions are found in cells, and sea water can be regarded as a complex electrolyte solution. In the past, aqueous solutions of electrolytes have been extensively studied.[1-5] Many properties have been analyzed in detail, for instance, densities,[6,7] viscosities,[6,8-10] diffusion coefficients,[11-14] conductivities,[6,8,14-16] and interfacial properties[17-19] for different salts at different conditions (temperature, pressure, or concentration). Even more complex properties, such as the hydrogen bonding structure in aqueous electrolyte solutions, have been the object of study.[20-22] These properties were considered in both experimental and theoretical studies. However, the advent of computer simulations in the 1950's made possible to study electrolytes using a new tool. After considering simple systems as hard spheres or noble gases[23] (described by the Lennard-Jones (LJ) potential), it was only in the 1970's that the first Molecular Dynamics (MD) studies of ionic systems were reported in the pioneering works of Heinzinger, Vogel, Singer, and Sangster.[24-28] The main drawback of computer simulations is that the interactions between molecules are not known exactly, and it is necessary to describe them in an approximate way usually denoted as the force field. In the case of aqueous electrolyte solutions, a force field for water and another one to describe ion–water and ion–ion interactions are needed.[29]

Concerning water, the first potential model of water was proposed in 1933 by Bernal and Fowler.[29] Later, in the 1980's, Jorgensen and co-workers developed several models (that are still widely used today), such as TIP3P,[30] TIP4P,[30] and TIP5P,[31] also followed by the popular SPC/E<!--Q1: Please define SPC/E at first occurrence.--> of Berendsen et al.[32] The combination of the TIP4P geometry and the polarization correction introduced by Berendsen et al.[32] led to a new generation of TIP4P models, for instance, TIP4P-Ew[33] and TIP4P/2005.[34] It has been shown that the TIP4P/2005 model reproduces a wide variety of properties of water, such as density, viscosity, diffusion coefficients, surface tension, or the temperature of the maximum in density (TMD).[35] Although no empirical potential is still able to describe all properties of pure water, we have at least some reasonable models.

In the case of electrolytes, in recent years, many force fields for both alkali and alkali-earth halide salts have been developed.[36-65] Target properties used to develop these force fields (with some exceptions) were limited to hydration free energies of the solvated ions, ion–water distances, or densities of the solutions at low concentrations. In some cases, several properties of the solid were considered as lattice energies and/or lattice constants. Let us briefly mention two popular force fields for ions in water. The first one (widely used at the present time) is that proposed by Joung and Cheatham,[50] who proposed a force field for alkali halides. They developed parameters for the ions to be used with different models of water (TIP3P, TIP4P-Ew, and SPC/E). Another popular force field is the one proposed by Smith and Dang,[42] which was developed in combination with SPC/E water.[32] This force field was also adjusted regarding the gas phase binding enthalpies for small ion–water clusters and solvation enthalpies of ionic solutions. Strikingly, these force fields did not use densities (at high concentrations), transport properties, solubilities, surface tensions, freezing depression, or temperature of maximum density as target properties.

Let us now briefly describe why these properties are of interest.

- Viscosities and diffusion coefficients are transport properties of high practical interest when considering electrolyte solutions. Kim *et al.*[12] showed that at high concentrations all force fields overestimate the decrease in the diffusion coefficient of water due to the addition of salt. The diffusion coefficient of water can be related (approximately) to the viscosity according to the Stokes–Einstein relation.[66] Thus, low values in the diffusion of water imply too high viscosities. In fact, Yue and Panagiotopoulos[67] and ourselves[65,68] demonstrated that this is indeed the case.

- Another interesting property is the liquid–vapor surface tension. It is well known that adding salt to water increases its surface tension. However, common force fields for electrolytes are not able to quantitatively reproduce this change. Most of them overestimate that change, and the Madrid model underestimates it slightly.[19,69]

- The solubility of a salt in water is another interesting property. However, evaluating the solubility of a certain force field is both difficult and costly from a computational point of view, and for these reasons, it has been studied in detail only in the recent years.[70-74] Recently, Tanaka *et al.*[64] have developed one of the first force fields (for NaCl and KCl) that are able to reproduce the experimental solubilities of NaCl and KCl when used in combination with the TIP3P, TIP4P/2005, and SPC/E models of water. In addition, Moucka *et al.*[75] developed a promising polarizable force field with reasonable solubilities. However, since solubility was never considered when developing force fields, it has been found that most of the force fields significantly underestimate the solubility, as it has been shown by Nezbeda *et al.*[76] and Panagiotopoulos.[77] Due to the low solubility of most of the force fields, it has been found that the number of contact ion pairs (CIPs), (i.e., a cation in contact with an anion in solution) was quite high[78-81] and aggregation of ions (which can be regarded as the initial step of precipitation) was observed in many simulations even at concentrations well below the experimental solubility.[82] Actually, ion clustering has been reported for different salts below its experimental solubility limit. This fact can be seen in monovalent salts as NaCl,[48,83-85] and KCl,[86] divalent salts as $CaCl_2$,[87] or even sulfates as $Na_2SO_4$[88] and $Li_2SO_4$.[89] Of course, one cannot obtain meaningful physical conclusions about electrolyte solutions in models where ions aggregate well below the value of the experimental solubility.

· Water has a maximum in density when cooled at constant pressure. The temperature at which the maximum in density occurs is usually denoted as the TMD. This fingerprint property of water is strongly related to its singular behavior.[90] When salt is added to water, the temperature at which this maximum in density occurs is shifted to lower temperatures, and even for a 1 m solution, the shift ranges from the value of 4° (for LiCl) to the value of 19° (for CsI).[91] If a model of electrolyte in water aims to describe how ions disrupt the structure of water, it should reproduce the experimental value of this shift. The TMD is not just another property of water, but it is probably one of the most important ones.

· Finally, when adding salt to water, the freezing point of water decreases. Reproducing this freezing point depression is also interesting as it reflects how the presence of ions affects the chemical potential of water. Since this property can be determined at the present time in computer simulations, it seems of interest to consider it.

Right now, there is no force field able to reproduce all these properties. There are several routes that could eventually lead to an entirely satisfactory description of intermolecular forces in ionic solutions.

One option is by means of quantum calculations.[92-96] However, the computational cost is high, the number of molecules simulated is typically quite small, and also density functional theory is not free of approximations. Another possibility is to use polarizable models. Kiss and Baranyai already proposed these types of models,[60,97] and we expect further progress along this path. However, the introduction of polarizability in the models is not synonymous of a correct description of all properties of the solution. For instance, the solubility of some polarizable models is still low.[74] Polarizable models are typically ten times more expensive than the not polarizable ones, and depending on the problem, that may exceed current computational power. However, some promising force fields have been proposed recently.[75,98]

Nevertheless, there is a cost-effective way of introducing some type of polarization and/or transfer of charge: simply scaling the charges of the ions in the force field. In these scaled-charge force fields, the charge of monovalent ions is not ±1 but is reduced to a charge smaller than one. The use of scaled charges in simulations of electrolyte solutions started with Leontyev and Stuchebrukhov[99-104] who pointed out that the dielectric constant of non-polarizable models at high frequencies ($\epsilon_\infty$) was 1, and for water, the value was 1.78. They proposed a solution for this, and it was the use of a scaled charge of ± 0.75 (in electron units), which comes from applying a factor $q_{scaled} = 1./\sqrt{(\epsilon_\infty)}$ to the ions [this is also denoted as the Electronic Continuum Correction (ECC)]. As will be shown in this work, this value of the scaled charge is not able to reproduce all properties of electrolytes in water, but it turns out to be the optimum value to describe transport properties. There is another way of thinking proposed by Kann and Skinner,[105] which also leads to a scaled charge for the ions. In this case, they suggested that the Coulombic energy between ions at infinite dilution and infinitely large distances should be the same for the experiment and for the simulation. In this way, Debye–Huckel law would be recovered. If one uses this approach, the charge of the ions depends on the value of the dielectric constant of the model. In the particular case of the TIP4P/2005 force field, the dielectric constant is about 57, and this leads to a value of ±0.85 for the ions. Another argument on the use of scaled charges is just to recognize that these charges are just fitting parameters, and one can use different parameters to describe the potential energy surface (PES) and the dipole moment surface (DMS)[106,107] as it is often done when implementing neural networks for both surfaces.[108] Although integer charges should probably be used to describe the DMS, scaled charges could provide a better description of the PES.[106,109] During these last years, the use of scaled charges has been growing. Jungwirth and co-workers,[87,89,110-112] following the route of Leontyev and Stuchebrukhov, proposed a force field for several salts with a charge of ±0.75 for the ions in combination with SPC/E or TIP4P/2005 water. Higher charges (±0.885) were proposed for NaCl[113] and KBr[114] in combination with TIP4P/$\epsilon$ water[115] by Barbosa and co-workers or Li and Wang[116] with charges close to ±0.80 employing water BLYPSP-4F.[117] Bruce and van der Vegt[118] also investigated the use of scaled charges for trivalent salts in combination with SPC/E water. Some authors have gone even further by proposing that the charge transfer between ions and water is different for the cation and for the anion. This is the case of the work of Rick and co-workers[119-122] and Yao *et al.*[123] who suggested that the charges of cations should be around +0.9 and those of anions should be around −0.75 to −0.8. The surrounding water molecules would remain charged in order to maintain the electro-neutrality of the system. It is also true that if one uses a water model, such as mW,[124] which has no charges, it is possible to develop models for ions without charges, mimicking ionic effects with short-ranged interactions and obtaining reasonable results.[125] In any case, it seems that the use of scaled charges in electrolyte solutions is gaining relevance as Jungwirth has recently summarized in a couple of reviews in which the advantages of using scaled charges were discussed.[126,127]

Keeping this in mind, in 2017, we developed a force field for NaCl[69] with scaled charges (±0.85) in combination with TIP4P/2005 water.[34] Later, in 2019, we developed a new force field[68] with scaled charges for several salts followed up by an extension to a larger set of salts.[65] This force field has been denoted as the Madrid-2019 force field. This force field provides a correct description of the properties of seawater,[128] the TMD of different salts,[91] the depression of the freezing temperature,[129] or the salting out of methane in electrolyte solutions.[130] Scaled charges also improve (although the agreement was not quantitative yet) the results for transport properties (diffusion coefficients or viscosities) when compared to models using formal charges.[67]

The goal of this work is rather simple. A simple electrolyte solution will be chosen: NaCl in water. Several "Madrid"-like models will be developed using TIP4P/2005 for water and different values of the scaled charge for the ions. The purpose is to analyze whether there is a single value of the scaled charge able to describe the experimental values for all the properties considered in this work. As will be shown, this is not the case. However, we have found that certain values of the scaled charge predict extraordinary well certain properties. Finally, we shall consider the case of KCl in water. We shall show that the conclusions obtained for NaCl also hold for KCl. That strongly suggests that the conclusions of this work seem to be general and most likely also apply to other salts.

## II. FORCE FIELDS WITH DIFFERENT SCALED CHARGES

In this work, water interactions will be described by using the TIP4P/2005 model of water. For pure water, this model provides a good description of all properties considered in this work: structure, densities, TMD, transport properties (diffusion coefficient and viscosity), and surface tension.[35] The model uses the TIP4P geometry (already proposed by Bernal and Fowler[29] and used by Jorgensen *et al.*[30] when developing the TIP4P model). The parameters of this water model are provided in Table I.

TABLE I. Force field parameters for water TIP4P/2005.[34]

| Molecule | Charge ($e$) | $\sigma_{ii}$ (Å) | $\epsilon_{ii}$ (kJ/mol) |
|---|---|---|---|
| TIP4P/2005 | | | |
| O | 0 | 3.1589 | 0.7749 |
| H | 0.5564 | | |
| M | −1.1128 | | |

We shall use non-polarizable force fields that describe the interactions between ions and between ions and water by a Lennard-Jones (LJ) potential plus Coulombic interactions. Scaled charges will be used for the ions Na$^+$ and Cl$^-$. We shall always use electron units when referring to scaled charges. For the particular case of ±0.85, we shall use the parameters of the Madrid-2019 force field. In this work, we shall consider other possible values of the scaled charge. In particular, we shall consider ±0.75, ±0.80, and ±0.92. For each value of the scaled charge, we shall develop a new force field by fitting the parameters of the LJ interactions for ion–ion and ion–water interactions. It should be recognized that we cannot provide a physical interpretation/meaning to each individual value of the scaled charge. We rather consider the value of the scaled charge as an additional adjustable parameter of the force field. One could use, in principle, scaled charges larger than one, in general, that deteriorates the agreement with the experimental results for all properties analyzed in this work. For that reason, charges with |q| > 1 will not be considered in this work. The protocol used to develop all force fields of NaCl obtained in this work was always the same. First, we set the value of the charge for the ions and then we adjusted the LJ parameters for the ion–water interactions to correctly reproduce the densities over the whole range of concentrations up to the solubility limit of each salt. Then, we adjusted the cation–anion interaction in order to keep under control (i.e., smaller than 0.5) the number of contact ion pairs, avoiding the precipitation of the salt. We have shown that for many force fields, the number of CIPs at the solubility limit of the model is smaller than 0.5.[131] It seems reasonable then to impose this constraint at the experimental values of the solubility to avoid any clustering of ions and precipitation. The values of the LJ parameters for cation–cation and anion–anion interactions were kept constant for all values of the scaled charge. These parameters affect very little the properties of the solution up to the solubility limit of NaCl. However, they should be adjusted if one is interested in describing the melt and/or the solid phase of NaCl. Using scaled charges to describe the melt is probably not a good idea,[132] and one should limit the use of scaled charges to aqueous solutions. For the case of models with formal charges for the ions (i.e., ±1), we shall consider two force fields. The first one is the model proposed by Yagasaki et al.[64] that was specially designed for TIP4P/2005 and that reproduces the experimental value of the solubility of NaCl. The second one will be denoted as JC-TIP4P/2005 and was first proposed by Benavides et al.[69] In this force field, the parameters of NaCl proposed by Joung and Cheatham (for SPC/E) are combined with TIP4P/2005 water and use Lorenz–Berthelot combining rules for the cross interactions.

Let us now present the parameters of the different models developed in this work. In Table II, we show the parameters obtained for the models with charges ±0.75, ±0.80, ±0.85 (Madrid-2019), and ±0.92. All these models will be labeled as Madrid, but we shall add a word that emphasizes the property for which the model provides a better description of the experimental results for NaCl.

TABLE II. Force field parameters for NaCl models with different charges developed in this work and for Madrid-2019.[68] $\sigma_{iOw}$ and $\epsilon_{iOw}$ are the cross interactions between the water and ions.

| Model | Charge (e) | $\sigma_{ii}$ (Å) | $\epsilon_{ii}$ (kJ/mol) | $\sigma_{iOw}$ (Å) | $\epsilon_{iOw}$ (kJ/mol) | $\sigma_{Na-Cl}$ (Å) | $\epsilon_{Na-Cl}$ (kJ/mol) |
|---|---|---|---|---|---|---|---|
| Model q = 0.92 (Madrid-Interfacial) | | | | | | | |
| Na | 0.92 | 2.217 37 | 1.472 356 | 2.757 375 4 | 0.793 388 | 3.183 123 1 | 1.438 894 |
| Cl | −0.92 | 4.699 06 | 0.076 923 | 4.279 669 8 | 0.061 983 | | |
| Model q = 0.85 (Madrid-2019) | | | | | | | |
| Na | 0.85 | 2.217 37 | 1.472 356 | 2.608 38 | 0.793 388 | 3.005 12 | 1.438 894 |
| Cl | −0.85 | 4.699 06 | 0.076 923 | 4.238 67 | 0.061 983 | | |
| Model q = 0.80 | | | | | | | |
| Na | 0.80 | 2.217 37 | 1.472 356 | 2.493 6 | 0.793 388 | 2.880 12 | 1.438 894 |
| Cl | −0.80 | 4.699 06 | 0.076 923 | 4.188 01 | 0.061 983 | | |

| Model q = 0.75 (Madrid-Transport) | | | | | | | |
|---|---|---|---|---|---|---|---|
| Na | 0.75 | 2.217 37 | 1.472 356 | 2.387 25 | 0.793 388 | 2.580 12 | 1.438 894 |
| Cl | −0.75 | 4.699 06 | 0.076 923 | 4.076 31 | 0.061 983 | | |

In Fig. 1, densities of NaCl solutions for all "Madrid" force fields of this work are presented. Results were obtained at 298.15 K and 1 bar. Note that the results for q = ±0.80, ±0.85, and ±0.92 were shifted up 100, 200, and 300 kg/m$^3$, respectively, for a better legibility. It can be seen that for all values of the scaled charge, the experimental densities of the NaCl solutions are well reproduced. Although this is somewhat expected as we used the experimental density at low-moderate concentration as a target property, it was not obvious that densities could be reproduced in the whole concentration range. Thus, density is not a property that allows us to discriminate, which is the best value of the scaled charge that one should use as it is possible to reproduce the experimental densities with several scaled charges (while in the range ±0.75 - ±1). In any case, in our opinion, the experimental densities should always be used as a target property when developing force fields. Firstly, because their experimental values are known with extremely high accuracy (which is not the case, for instance, for the ion–oxygen distance, which is known indirectly and with high uncertainty[133,134]) and, secondly, because when the density is predicted correctly, deviations between the model predictions and the experiment indicate without any ambiguity deficiencies in the force field.

FIG. 1. Density as a function of molality at T = 298.15 K and 1 bar for NaCl aqueous solutions. Blue circles indicate the results of this work for the different developed force fields and Madrid-2019. Black solid lines indicate the fit of experimental data taken from Ref. 7. Results for q = ± 0.80, ±0.85, and ±0.92 were shifted up 100, 200, and 300 density units, respectively, for a better legibility.

In Table III, the Na–O$_w$ and Cl–O$_w$ distances (being O$_w$ the oxygen of water), the number of contact ion pairs (CIPs), and the hydration numbers of Na and Cl are presented. The number of contact ion pairs (CIPs) can be evaluated from the radial cation–anion distribution function using the following equation:

$$n^{CIP} = 4\pi\rho_{\pm} \int_{0}^{r_{\min}} g_{+-}(r)\, r^2\, dr, \tag{1}$$

where $g_{+-}$ is the cation–anion radial distribution function (RDF) and $\rho_{\pm}$ is the number density of cations or anions (number of cations/anions per unit of volume) and $r_{\min}$ (the integral upper limit) is the position of the first minimum in the RDF, which must be located at a similar distance to that of the cation–$O_w$ RDF. A simultaneous plot of the cation–anion and cation–$O_w$ RDFs is useful to determine if one is really evaluating CIP or a solvent separated ion pair (SIP). To evaluate the hydration numbers, the procedure is the same but using the ion–$O_w$ RDF instead. As can be seen, there is a clear trend in the ion–water distances. With respect to the cation (i.e., $Na^+$), it is clear that as the value of the scaled charge increases, the Na–$O_w$ distance (i.e., distance of the first peak in the RDF) increases. The same trend can be observed with the hydration number of the cation. In the case of the anion, the trend is the same, but the differences in the anion–$O_w$ distances are smaller than in the case of the cation. Finally, as the models have been developed following a similar strategy, the number of CIPs is similar and is rather low (which guarantees the absence of precipitation of the salts).

TABLE III. Structural properties for NaCl aqueous solutions evaluated with different force fields at 298.15 K and 1 bar. Number of contact ions pairs (CIPs), hydration number of cations ($HN_c$) and anions ($HN_a$), and position of the first maximum of the cation–water $(d_{c-O_w})$, and anion–water $(d_{a-O_w})$ in the radial distribution function. Experimental data are taken from the works of Dang et al.[135] and Tongraar et al.[136] for $Cl^-$ and from the work of Galib et al.[137] for $Na^+$. Results in bold indicate significant deviation from the experimental values.

| Charge (*e*) | Model | *m* (mol/kg) | CIP | HN$_c$ | HN$_a$ | $d_{c-O_w}$ (Å) | $d_{a-O_w}$ (Å) |
|---|---|---|---|---|---|---|---|
| q = ±0.75 | Madrid-Transport | 1 | 0.05 | 4.4 | 5.3 | **2.12** | **2.94** |
| q = ±0.80 | | 1 | 0.02 | 5.1 | 5.5 | **2.23** | 3.02 |
| q = ±0.85 | Madrid-2019 | 1 | 0.03 | 5.5 | 5.9 | 2.32 | 3.05 |
| q = ±0.92 | Madrid-Interfacial | 1 | 0.03 | 5.9 | 6.3 | 2.46 | 3.05 |
| q = ±1 | JC-TIP4P/2005 | 1 | 0.02 | 6.0 | 6.8 | 2.41 | 3.13 |
| q = ±1 | Yagasaki model | 1 | 0.01 | 5.9 | 6.9 | 2.36 | 3.16 |
| | Experimental[135,137] | | | 5.2–5.8 | 5.4–7.4 | 2.35–2.39 | 3.08–3.14 |
| | Experimental[136] | | | | 4.3–7.7 | | 3.01–3.09 |

The models with q = ±1, q = ±0.92, and q = ±0.85 describe well the experimental ion–O$_w$ distances and hydration numbers. For models with lower charge (i.e., q = ±0.8 and specially q = ±0.75), deviations from the experimental values are clearly seen (they are presented in bold in Table III). It should be reminded that these distances (and the corresponding hydration numbers) present large uncertainties as they are obtained indirectly from diffraction experiments.[133] It is somewhat surprising that it is possible to reproduce the experimental densities in models with different values of the ion–O$_w$ distance. In principle, one would expect lower densities in models with larger ion–O$_w$ distance as the volume excluded to water by the ions is larger in this case. However, it is clear that this is only half of the story as this is compensated by a more effective packing of water beyond the first hydration shell of the ions in this case (i.e., the hydration bonded network is more disrupted). In Sec. III, we will describe the numerical details of simulations that will be carried out to determine the rest of the properties of this work.

# III. SIMULATION DETAILS

We have performed Molecular Dynamics (MD) simulations with GROMACS[138,139] in the NpT and NVT ensembles. The leap-frog integrator algorithm,[140] with a time step of 2 fs, has been employed in all simulations. Periodic boundary conditions in all directions were also applied in all runs. The temperature and pressure were kept constant using the Nosé–Hoover thermostat[141,142] with a coupling constant of 2 ps for temperature and the Parrinello–Rahman barostat[143] with a time constant of 2 ps for pressure. For electrostatics and van der Waals interactions, the cut-off radii was fixed at 1.0 nm and long-range corrections in the energy and pressure were applied to the Lennard-Jones part of the potential. The smooth PME<!--Q3: Please define PME at first occurrence.--> method[144] to account for the long-range electrostatic forces was used. Water geometry was maintained using the LINCS algorithm.[145,146] The densities of our models have been evaluated with NpT simulations of 50 ns for a system containing 555 water molecules with the corresponding number of ions to reproduce the desired concentration. With this number of molecules of water, a 1 m solution is obtained by using ten molecules of NaCl. Concentrations will be given in this work in molality units (i.e., number of moles of salt per kilogram of water).

For the calculation of the transport properties (i.e., viscosities and diffusion coefficients), we have used a large system containing 4440 molecules of water (555 × 8 = 4440) and the corresponding number of ions (10 × 8 = 80 for 1 m concentration). To evaluate the viscosities, we have followed the methodology proposed by González and Abascal.[147] A previous NpT simulation of 20 ns is carried out to calculate accurately the volume of the system. After that, we perform a NVT simulation of 50 ns using the average volume obtained in the NPT simulation. The pressure tensor $p_{\alpha\beta}$ was calculated and saved on disk every 2 fs. Finally, we used the Green–Kubo formula for the shear viscosity $\eta$.

$$\eta = \frac{V}{kT} \int_{0}^{\infty} \langle p_{\alpha\beta}(t_{0})\, p_{\alpha\beta}(t_{0}+t) \rangle_{t_{0}}\, dt, \quad (2)$$

where *V* is the volume of the system, *k* is the Boltzmann constant, T is the temperature, and $p_{\alpha\beta}$ are the non-diagonal components of the pressure tensor. The upper limit of the integral is usually between 10 and 20 ps. Diffusion coefficients of water in NaCl aqueous solutions have been calculated using the system of 4440 water molecules, which was used to evaluate the viscosity. The Einstein relation was used to calculate diffusion coefficients:

$$D_{MD} = \lim_{t \to \infty} \frac{1}{6t} \left\langle [\vec{r}_{i}(t) - \vec{r}_{i}(t_{0})]^{2} \right\rangle, \quad (3)$$

where $\vec{r}_{i}(t)$ and $\vec{r}_{i}(t_{0})$ are the position of the i-th particle at time t and a certain origin of time $t_0$ and the $\langle [r_{i}(t) - r_{i}(t_{0})]^{2} \rangle$ term is the mean square displacement (MSD). For all the results of this work, we applied the hydrodynamic corrections of Yeh and Hummer,[148] which are described as follows:

$$D = D_{MD} + \frac{kT\xi}{6\pi \eta L} \quad (4)$$

where *D* is the diffusion coefficient with the applied corrections of Yeh and Hummer, $D_{MD}$ is the diffusion coefficient initially obtained by simulations, $\xi$ is a constant (its value is 2.837), $\eta$ is the viscosity of the model at the studied concentration, and L is the length of the simulation box.

Surface tension has been evaluated by using the direct coexistence method. We have performed NVT simulations of 40 ns employing a system of 6660 water molecules and the corresponding number of ions (120 for a 1 m solution) and using a cutoff of 1.4 nm (the surface tension is quite sensitive to the value selected for the cutoff). In the initial configuration, the solution occupied about one third of the simulation box, and the rest was filled with vapor. In this case, no long range corrections to the LJ part of the potential were used. The surface tension of each model is calculated as usual using the following expression:

$$\gamma = \frac{L_z}{2}(p_{zz} - (p_{xx}+p_{yy})/2) \quad (5)$$

where $p_{zz}$ is the normal component of the pressure and $p_{yy}$ and $p_{xx}$ are the tangential components of the pressure (being the z axis perpendicular to the liquid–vapor interface).

For the calculation of the freezing depression, the methodology followed is based on the direct coexistence of two phases: a solid phase consisting of ice Ih (2000 ice molecules) and a liquid phase consisting of an aqueous NaCl solution of a given concentration (2000 water molecules). The ice plane exposed at the interface is the secondary prismatic one (1$\bar{2}$10) following the same approach as Conde et al.[149] Finally, for evaluating the temperature of maximum density (TMD), we have used systems of 555 water molecules and ten molecules of NaCl (i.e., 1 m solutions). The simulation length in this case is about 150 ns, and we typically selected six or seven temperatures along the room pressure isobar.

Let us finish by mentioning that for all force fields considered in this work, precipitation did not occur at the highest considered concentration (i.e., the solubility limit which is 6.15 m for NaCl and 4.81 m for KCl) even after very long runs. This is important to guarantee that the results of this work are not affected by the appearance of solid clusters (as often found in previous studies with integer charges when studied at the experimental solubility limit).

## IV. RESULTS

### A. Viscosities of aqueous NaCl solutions

Let us begin by presenting the results for the viscosities of aqueous NaCl solutions by employing the different force fields of this work. In previous studies,[65,68] it was shown that using a scaled charge of ±0.85 (Madrid-2019) improved the description of the viscosities when compared to the results obtained using formal charges. Nevertheless, even the Madrid-2019 model was not able to reproduce the experimental viscosities specially at very high concentrations.

In Fig. 2, we show the results for the viscosities of NaCl solutions evaluated with different force fields. We first look at the results of a unit charge model. The JC-TIP4P/2005 model significantly overestimates the viscosity of the aqueous NaCl solutions. Deviations are clearly visible even at 1 m. Decreasing the charge to ±0.92 reduces the viscosity by a large amount, but the results are still far away from the experimental line. The Madrid-2019 force field improves the predictions with respect to the two previous models. However, deviations from experiment are still evident for concentrations above 2 m. Our next model, with a charge of q = ±0.80, improves the results of Madrid-2019. Further reducing the charge to ±0.75 allows us to obtain quantitative agreement with the experiment. At last, we are able to reproduce the experimental viscosities of NaCl aqueous solutions in the whole concentration range. Thus, we conclude that the viscosity is strongly affected by the charge of the ions (see the numerical results in Table IV). The only charge able to reproduce experimental viscosities of the NaCl aqueous solution for the whole range of molalities is q = ±0.75. For this reason, this force field will be denoted as Madrid-Transport. It should mentioned that also for ionic liquids scaled charges improve dramatically the description of transport properties.[150,151]

FIG. 2. Shear viscosity curves as a function of concentration for aqueous NaCl solutions at 298.15 K and 1 bar. Models studied in this work: q = ±1 JC-TIP4P/2005 (magenta triangles), q = ±0.92 Madrid-Interfacial (orange diamonds), q = ±0.85 Madrid-2019 (red squares), q = ±0.80 (cyan triangles), and q = ±0.75 Madrid-Transport (blue circles). The continuous line is the fit of the experimental data taken from Refs. 152 and 153.

TABLE IV. Results for viscosity obtained with the different models proposed in this work for NaCl solutions in TIP4P/2005 water at temperature T = 298.15 K and pressure p = 1 bar for different concentrations below experimental solubility. Experimental data were taken from Refs. 152 and 153.

| Molality (mol/kg) | Viscosity (mPa·s) | | | | | |
|---|---|---|---|---|---|---|
| m | Expt. | q = ±1 JC-TIP4P/2005 | q = ±0.92 Madrid-Interfacial | q = ±0.85 Madrid-2019 | q = ±0.80 | q = ±0.75 Madrid-Transport |
| 1 | 0.97 | 1.14 | 1.06 | 1.05 | 1.00 | 0.97 |
| 2 | 1.08 | 1.55 | 1.30 | 1.27 | 1.14 | 1.12 |
| 4 | 1.35 | 3.50 | 2.10 | 1.75 | 1.50 | 1.44 |
| 6 | 1.75 | 5.40 | 3.10 | 2.56 | 1.96 | 1.79 |

## B. Diffusion coefficients in aqueous NaCl solutions

Let us now consider the diffusion coefficients of water in the NaCl solutions. Since the Stokes–Einstein relation[66] relates the viscosity and the diffusion coefficient, one may expect that a good description of the viscosity implies a good description of the diffusion coefficient. In Fig. 3, the diffusion coefficients of water in NaCl solutions at different concentrations are shown, and in Table V, the numerical results are collected. In all cases, we applied the Yeh and Hummer[148] finite size corrections (using the calculated viscosities of the models at different concentrations). The JC-TIP4P/2005 force field that uses formal charges highly underestimates the diffusion coefficient of water when adding salt. In fact, the change with concentration is quite different from the experimental one (the slope, in absolute value, is quite large). When decreasing the value of the scaled charge, we can observe a progressive improvement in the results, being the Madrid-Transport force field (q = ±0.75) the one with the best agreement with the experimental results. As in the case of the viscosity, it is clear that the decrease of the charge of the ions leads to a better description of the diffusion coefficients. To the best of our knowledge, this is the first time a model of NaCl in water is able to reproduce the experimental values of viscosities and of the diffusion coefficients of water up to high concentrations, thus overcoming the challenge raised in the year 2012 by Kim et al.[12]

FIG. 3. Diffusion coefficients of water in NaCl solutions (at 298.15 K and 1 bar) at different concentrations. The results include hydrodynamic corrections of Yeh and Hummer.[148] Magenta triangles: results for the q = ±1 JC-TIP4P/2005 force field. Orange diamonds: results for q = ±0.92 Madrid-Interfacial. Red squares: results for the q = ±0.85 Madrid-2019 model. Cyan triangles: q = ±0.80 model. Blue circles: results for the q = ±0.75 Madrid-Transport model. The continuous line is the fit of the experimental data taken from Ref. 11.

TABLE V. Results for the diffusion coefficients of water obtained with the different models studied in this work for NaCl solutions in TIP4P/2005 water at temperature T = 298.15 K and pressure p = 1 bar for different concentrations below experimental solubility. The results include hydrodynamic corrections of Yeh and Hummer.[148] Expt. data were taken from Ref. 11.

| Molality (mol/kg) | $D \cdot 10^5$ ($cm^2/s$) |
| --- | --- |

| m | Expt. | q = ±1 | q = ±0.92 | q = ±0.85 | q = ±0.80 | q = ±0.75 |
|---|---|---|---|---|---|---|
| | | JC-TIP4P/2005 | Madrid-Interfacial | Madrid-2019 | | Madrid-Transport |
| 1 | 2.17 | 1.85 | 1.95 | 2.02 | 2.06 | 2.09 |
| 2 | 2.02 | 1.45 | 1.65 | 1.72 | 1.84 | 1.89 |
| 4 | 1.71 | 0.86 | 1.15 | 1.29 | 1.45 | 1.58 |
| 6 | 1.42 | 0.50 | 0.80 | 0.94 | 1.14 | 1.31 |

We shall now discuss the values of the diffusion coefficients of the individual ions (i.e., $Na^+$ and $Cl^-$). In Tables VI and VII, we have collected the diffusion coefficient of $Na^+$ and $Cl^-$ for the different studied force fields and at different concentrations. Regarding the experimental values, the diffusion coefficient of $Cl^-$ at infinite dilution is higher than the diffusion coefficient of $Na^+$. This trend is captured by all force fields (i.e., higher values for the self-diffusion coefficient of $Cl^-$ than for $Na^+$). It is also interesting to study the change of the diffusion coefficient of the ion when increasing the salt concentration. All models follow the same behavior [as we show in Figs. 4(a) and 4(b)]; the diffusion coefficient of the ions decreases when increasing the salt concentration. Finally, the impact of the charge on the diffusion coefficient is clear: the diffusion coefficient of both $Na^+$ and $Cl^-$ increases when the charge decreases. Again, the model with q = ±0.75 (Madrid-Transport) is the one that better reproduces the experimental diffusion coefficients of the ions. In the future, it would be of interest to study whether the model with q = ±0.75 also reproduces other transport properties, such as the electrical conductivities. Concerning dielectric properties in the supplementary material, we present results for the relative change of the dielectric constant of NaCl solutions for different force fields developed in this work. We have found that for charge q = ±1, the relative change is overestimated, for q = ±0.75, it is underestimated, and for q = ±0.92, we found excellent agreement with experimental data.

TABLE VI. Results for the diffusion coefficients of $Na^+$ cation obtained with the different models studied in this work for NaCl solutions in TIP4P/2005 water at temperature T = 298.15 K and pressure p = 1 bar for different concentrations below experimental solubility. The results include hydrodynamic corrections of Yeh and Hummer.[148] Experimental results at infinite dilution have been taken from Ref. 154.

| Molality (mol/kg) | $D \cdot 10^5$ $Na^+$ (cm$^2$/s) | | | | |
|---|---|---|---|---|---|
| m | q = ±1 | q = ±0.92 | q = ±0.85 | q = ±0.80 | q = ±0.75 |
|   | JC-TIP4P/2005 | Madrid-Interfacial | Madrid-2019 |   | Madrid-Transport |
| 0 (Expt.) | 1.33 | 1.33 | 1.33 | 1.33 | 1.33 |
| 1 | 0.80 | 1.04 | 1.11 | 1.24 | 1.31 |
| 2 | 0.64 | 0.90 | 0.96 | 1.10 | 1.19 |
| 4 | 0.37 | 0.63 | 0.73 | 0.89 | 0.97 |
| 6 | 0.23 | 0.45 | 0.54 | 0.70 | 0.82 |

TABLE VII. Results for the diffusion coefficients of the $Cl^-$ anion obtained with the different models studied in this work for NaCl solutions in TIP4P/2005 water at temperature T = 298.15 K and pressure p = 1 bar for different concentrations below experimental solubility. The results include hydrodynamic corrections of Yeh and Hummer.[148] Experimental results at infinite dilution have been taken from Ref. 154.

| Molality (mol/kg) | $D \cdot 10^5$ $Cl^-$ (cm$^2$/s) | | | | |
|---|---|---|---|---|---|
| m | q = 1 | q = 0.92 | q = 0.85 | q = 0.80 | q = 0.75 |

|   | JC-TIP4P/2005 | Madrid-Interfacial | Madrid-2019 |  | Madrid-Transport |
|---|---|---|---|---|---|
| 0 (Expt.) | 2.03 | 2.03 | 2.03 | 2.03 | 2.03 |
| 1 | 1.13 | 1.24 | 1.45 | 1.51 | 1.59 |
| 2 | 0.88 | 1.05 | 1.21 | 1.35 | 1.43 |
| 4 | 0.51 | 0.74 | 0.90 | 1.07 | 1.17 |
| 6 | 0.29 | 0.50 | 0.66 | 0.85 | 0.95 |

FIG. 4. Diffusion coefficients of (a) $Na^+$ and (b) $Cl^-$ in NaCl solutions (at 298.15 K and 1 bar) at different concentrations. The results include hydrodynamic corrections of Yeh and Hummer.[148] Magenta triangles: results for the q = ±1 JC-TIP4P/2005 force field. Orange diamonds: results for q = ±0.92 Madrid-Interfacial. Red squares: results for the q = ±0.85 Madrid-2019 model. Cyan triangles: q = ±0.80 model. Blue circles: results for the q = ±0.75 Madrid-Transport model. Experimental results at infinite dilution (black crosses) were taken from Ref. 154.

## C. Surface Tension

We shall now consider the surface tension of the aqueous electrolyte solution when in contact with its vapor. The gas phase is practically pure water as the ions do not go into the gas phase. We shall present results for the change in the surface tension (i.e., $\Delta \gamma = \gamma_{NaCl\ solution} - \gamma_{pure\ water}$) versus the concentration to illustrate the impact of the ions in the surface tension. Note though that TIP4P/2005 provides a good estimate of the surface tension of pure water at room temperature[155,156] so that if the change in surface tension is predicted correctly, then the predictions for the absolute values of the surface tension will also be accurate. In Fig. 5, results for $\Delta \gamma$ as a function of the salt concentration by using the "Madrid" force fields are presented. The model with the formal charge (JC-TIP4P/2005) overestimates the increase in the surface tension of water due to the addition of the salt (in agreement with previous results[69]). The model with charge q = ±0.92 is in excellent agreement with the experimental change in surface tension in the whole concentration range, and for this reason, it will be labeled as Madrid-Interfacial. On the other hand, the model with the charge q = ±0.85 (Madrid-2019 model) slightly underestimates the increment of surface tension. Finally, the model with the charge q = ±0.75 (Madrid-Transport) highly underestimates the change in the surface tension due to the addition of salt. We have also calculated the surface tension for both pure water and a NaCl solution described by the q = ±0.92 Madrid-Interfacial model by using a much larger cutoff (i.e., 2.5 nm). We have found that values of the surface tension are about 2 mN·m$^{-1}$ higher when the cutoff is increased from 1.4 to 2.5 nm. However, the change in the surface tension in the solution with respect to pure water is not affected by the value of the cutoff as can be seen in Fig. 5.

Any hope that the model with charge q = ±0.75 that accurately described transport properties would also improve the description of the rest of properties is gone. The fact that the model with q = ±0.92 does a good job is maybe not so surprising. In the vapor phase, probably the value q = ±1 better describes the ions, whereas the value q = ±0.85 seems to be adequate for the liquid phase. In the interface, one has an "intermediate" situation and the charge q = ±0.92 should be regarded as a "mean field" value.

FIG. 5. Surface tension of NaCl aqueous solutions relative to that of pure water evaluated with identical simulation conditions at 298.15 K. Magenta line: results for the q = ±1 JC-TIP4P/2005 model. Orange line: results for the q = ±0.92 Madrid-Interfacial model. Green dashed line: results for the q = ±0.92 Madrid-Interfacial model evaluated with a cutoff of 2.5 nm. Red line: results for the q = ±0.85 Madrid-2019 model. Blue line: results for the q = ±0.75 Madrid-Transport model. Note that we have employed a cutoff of 1.4 nm for all cases except for the special case of Madrid-Interfacial in which we have also evaluated the surface tension with a cutoff of 2.5 nm. The estimated error for our Δγ results is about 0.8 mN·m$^{-1}$. The black symbols are the experimental results taken from Ref. 157.

## D. Freezing temperature depression

Adding salt to water decreases the freezing temperature of the solution. The difference between the freezing temperature of the solution and the freezing temperature of pure water is known as the freezing depression. The freezing depression of ices was studied first in the pioneering work of Kim and Yethiraj[158] and in 2018 by Conde et al.[149] with NaCl models with formal charges. They found that for low concentrations, the model worked properly, but as the concentration of NaCl increased, the results deviated from the experimental ones. Recently, Lamas et al.[129] have done a similar study by using the Madrid-2019 model and analyzing the behavior of several different salts, concluding that Madrid-2019 provides quite reasonable results although still with some room for improvement. Bearing this in mind, in this work, we decided to study the depression in the freezing temperature of an aqueous NaCl solution using the different models considered in this work differing in the value of the scaled charge.

To determine the freezing point depression, we follow the methodology of Lamas et al.[129] and we shall perform computer simulations of pure ice in contact with a NaCl aqueous solution. Following the phase rule, for a certain fixed pressure (i.e., room pressure) and temperature, the system will reach equilibrium for a certain value of the concentration of the salt in the aqueous solution. This equilibrium is reached either my melting some ice (thus decreasing the concentration of NaCl in the aqueous phase from the initial value) or by freezing some water (thus increasing the concentration of NaCl in the aqueous phase from its initial value). Not that the solubility of NaCl in ice is extremely low, and for this reason, we shall use pure ice (in the hexagonal phase) as the solid phase. For pure water, the melting temperature of ice Ih when using the TIP4P/2005 model is $T_m$ = 250 K (i.e., 23 K below the experimental value). We have decided to estimate the concentration of the NaCl aqueous solution when in equilibrium with ice at a temperature of 236 K. If we define $\Delta T = T - T_m$, this corresponds to a supercooling of $\Delta T$ = $-$14 K.

Figure 6 illustrates how the technique works. It can be seen for the model with q = $\pm$0.75 (Madrid-Transport) either starting from a solution with 3 m or from 4 m, one reaches the same equilibrium concentration of 3.6 m. This shows that equilibrium is reached regardless of the initial concentrations. However, reaching equilibrium is time consuming, and for this reason, it is more convenient for each value of the scaled charge to do some exploratory runs to obtain preliminary estimates of the value of the equilibrium concentration. After this value is estimated, we run a very long run starting from this value so that equilibrium is reached rapidly.

Figure 7 shows how the molality of the NaCl aqueous phase changes with time for the different models considered in this work. Initial values of the concentration were 3 m for the q = ±1 force fields (i.e., Yagasaki and JC-TIP4P/2005 models), 3.6 m for the model with q = ±0.75 (Madrid-Transport), and 4 m for the models with q = ±0.92 (Madrid-Interfacial) and q = ±0.85 (Madrid-2019). To calculate the equilibrium concentration, we have simulated all the models during 2 $\mu$s and we have averaged the molality of the last 1 $\mu$s.

FIG. 6. Molality of the aqueous solution phase as a function of time for the model with q = ±0.75 (Madrid-Transport) starting from different concentrations: 3 m (green circles), 3.6 m (blue circles), and 4 m (black circles) at 1 bar and 236 K (i.e., $\Delta$T = −14 K).

FIG. 7. Molality of the aqueous solution phase as a function of the simulation time for different models evaluated in this work at 1 bar and 236 K (i.e., $\Delta$T = −14 K). Red squares: Results for q = ±0.85 (Madrid-2019). Blue circles: q = ±0.75 (Madrid-Transport). Green up triangles: q = ±1 (Yagasaki Model). Pink down triangles: q = ±1 (JC-TIP4P/2005). Orange diamonds: q = ±0.92 (Madrid-Interfacial).

The equilibrium molalities at the studied $\Delta$T for the different models are represented in Fig. 8 and Table VIII. We can see that the unit charge models (JC-TIP4P/2005 and Yagasaki Model) with charges ±1.0 for the ions underestimate the equilibrium concentration (providing very similar results despite being different models). Models with q = ±0.92 (Madrid-Interfacial) and q = ±0.85 (Madrid-2019) overestimate the equilibrium concentration (they have higher molalities than the experimental one). Finally, the model with scaled charges ±0.75 (Madrid-Transport) accurately describes the experimental concentration at equilibrium for this supercooling. We do not find a regular behavior of the change of the freezing conditions with the value of the charge. It is important to point out that as discussed by Lamas et al.,[129] the freezing point depression is not only testing the force field for the electrolyte solution, but it is also testing the properties of ice (i.e., melting enthalpy and temperature) as they enter into the thermodynamic description, leading to the freezing point depression. The conclusion of this section is that the charge has an effect on the freezing depression of ice. However, trends with the value of the scaled charge are not monotonous, and at this point, we cannot find a correlation between the magnitude of the freezing point depression and the value of the scaled charges, pointing out that there must be several factors contributing to this. The summary is that deviations from the experimental value of the freezing point depression are moderate, and we do not see a clear correlation between the value of the scaled charge and the quality of the prediction (further work is required to understand that in detail).

TABLE VIII. Equilibrium concentrations of NaCl at $p = 1$ bar and $T = 236 K$ (i.e., supercooling of 14 K) when the NaCl aqueous solution is equilibrium with ice for the different force fields of this work. The reported experimental value is that obtained for a supercooling of 14 K.

| Charge ($e$) | Model | Initial concentration (mol/kg) | $\Delta T$ (K) | Equilibrium concentration (mol/kg) | Deviation from Expt. (mol/kg) |
|---|---|---|---|---|---|
| | Expt. | ... | 14 | 3.68 | 0 |
| $q = \pm 1$ | JC-TIP4P/2005 | 3 | 14 | 3.24 | 0.44 |
| $q = \pm 1$ | Yagasaki Model | 3 | 14 | 3.18 | 0.50 |
| $q = \pm 0.92$ | Madrid-Interfacial | 4 | 14 | 4.31 | −0.63 |
| $q = \pm 0.85$ | Madrid-2019 | 4 | 14 | 4.17 | −0.49 |
| $q = \pm 0.75$ | Madrid-Transport | 3.6 | 14 | 3.73 | −0.05 |

FIG. 8. Freezing point depression (at 1 bar) for the NaCl aqueous solution system evaluated in this work. Red squares: Results for $q = \pm 0.85$ (Madrid-2019) from the work of Lamas et al.[129] Cyan

cross: Result for q = ±0.85 (Madrid-2019) obtained in this work. Blue circle: q = ±0.75 (Madrid-Transport). Green up triangle: q = ±1 (Yagasaki model). Pink down triangle: q = ±1 (JC-TIP4P/2005). Orange diamonds: q = ±0.92 (Madrid-Interfacial). The red dashed line is the fit from Madrid-2019 results of Lamas et al.[129] The black continuous line is the fit of the experimental data taken from Refs. 159 and 160.

## E. Maximum in density of the electrolyte solution

We shall now examine the temperatures at which a maximum in density occurs for a NaCl solution with concentration 1 m. Recently, we have determined experimentally the TMDs for a large variety of salts and we have concluded that the Madrid-2019 model is able to accurately predict the TMDs.[91] It should be pointed out that the TIP4P/2005 model of water reproduces the experimental value of the TMD of pure water (i.e., 277 K). We think that TMD should be used as a target property when designing force fields both for pure water and for aqueous electrolyte solutions. We shall now analyze if the different force fields of this work are able to describe this property. In Fig. 9, the results for this property are shown. In Table IX, the values of the shift in the temperature of the TMD are also shown.

The trend is clear; models with a small value of the charge q = ±0.75, ±0.80 underestimate the shift in the TMD (and also the density at the maximum). Models with intermediate values q = ±0.85, ±0.92 describe extraordinary well the location of the TMD and the density at the maximum. Not that all these models describe quite well the experimental value of the density at room temperature as determined by Laliberte (open square in Fig. 9). However, it was not clear if they would be able to capture the subtle impact of ions into the structure of water when the temperature changes. It seems that models with q = ±0.85, ±0.92 capture this change quite well. The models with formal charges (JC-TIP4P/2005 and Yagasaki model) overestimate the shift in the TMD. They also overestimate the value of the density at the maximum although this is partly due to the fact that they also overestimate the experimental density at room temperature. The trend is clear; the larger the charge, the larger the change. Too small changes in the TMD suggest that the charge used is too small.

TABLE IX. Shift (at 1 bar) in the TMD (in K) with respect to pure water for the 1 m NaCl solutions studied in this work ($\Delta$ = TMD$_{Solution}$ - TMD$_{Water}$). The experimental TMD of pure water is 277.1 K, and for TIP4P/2005, it is 277.3 K.[91]

| Charge (e) | Model | $\Delta$ (K) |
|---|---|---|
|  | Expt. | −14.4 |
| q = ±1 | JC-TIP4P/2005 | −18.37 |
| q = ±1 | Yagasaki model | −19.2 |
| q = ±0.92 | Madrid-Interfacial | −15.3 |
| q = ±0.85 | Madrid-2019 | −16.1 |
| q = ±0.80 |  | −12.9 |
| q = ±0.75 | Madrid-Transport | −6.8 |

FIG. 9. Results (at 1 bar) for temperatures of the maximum density for 1 m NaCl solutions obtained with different salt models: q = ±1 Yagasaki model (green triangles), q = ±1 JC-TIP4P/2005 (magenta triangles), q = ±0.92 Madrid-Interfacial (orange diamonds), q = ±0.85 Madrid-2019 (red squares), q = ±0.80 (cyan triangles), and q = ±0.75 Madrid-Transport (blue circles). The black solid line is the fit of experimental data.[91] Values of densities at 298.15 K are shown as crosses. Experimental density at 298.15 K is shown as an black empty square.

The summary is that the force fields with charges q = ±0.85 and q = ±0.92 reproduce accurately the TMD and absolute densities of experimental aqueous NaCl solutions at 1 m and 1 bar.

## F. Transferability to other salts

To analyze whether the conclusions of this work could also be extended to other 1:1 electrolytes, we have also considered the case of KCl. We shall not perform an exhaustive study with all possible values of the scaled charge considered for NaCl (±1, ±0.92, ±0.85, ±0.80, and ±0.75). Rather, we shall only consider the cases, q = ±0.75, q = ±0.85, and q = ±0.92 for the scaled charge. For the cases q =

±0.75 and q = ±0.92 (KCl), we shall develop two new force fields following the procedure described before for NaCl. In this case, the LJ parameters of the anion Cl⁻ were identical to those used when developing the models of NaCl with q = ±0.75 and q = ±0.92. For the case q = ±0.85, we will take the parameters from the Madrid-2019 force field. In the case of q = ±1, we shall consider two force fields. The first one is that proposed by Yagasaki et al.[64] for KCl and TIP4P/2005 that reproduces the experimental value of the solubility of KCl. In the second one, we will combine the parameters of KCl proposed by Joung and Cheatham (for SPC/E) with TIP4P/2005 and use Lorenz–Berthelot combining rules for the cross interactions (i.e., the same approach that was used before for NaCl).

In Table X, we have collected the parameters for NaCl and KCl obtained in this work for the values of the scaled charges q = ±0.75 and q = ±0.92. In a sense, Table X constitutes the beginning of the Madrid-Transport and Madrid-Interfacial force fields. We expect to increase in the future the number of ions studied with these force fields.

TABLE X. Force field parameters for NaCl, KCl, NaOH, and KOH q = 0.75 Madrid-Transport and q = ±0.92 Madrid-Interfacial for KCl. LB means that Lorentz–Berthelot combining rules have been applied to the cross interactions. The bond length of O–H is 0.98 Å. Parameters for OH⁻ force field are taken from the work of Habibi et al.[161]

| Model | Charge (e) | $\sigma_{ii}$ (Å) | $\epsilon_{ii}$ (kJ/mol) | $\sigma_{iOw}$ (Å) | $\epsilon_{iOw}$ (kJ/mol) | $\sigma_{i\text{-}Cl}$ (Å) | $\epsilon_{i\text{-}Cl}$ (kJ/mol) |
|---|---|---|---|---|---|---|---|
| Madrid-Transport | | | | | | | |
| Na | 0.75 | 2.217-37 | 1.472-356 | 2.387-25 | 0.793-388 | 2.580-12 | 1.438-894 |
| K | 0.75 | 2.301-40 | 1.985-740 | 2.895-40 | 1.400-430 | 3.417-00 | 1.400-000 |
| Cl | −0.75 | 4.699-06 | 0.076-923 | 4.076-31 | 0.061-983 | | |
| OH | −0.75 | | | | | | |
| O(OH) | −1.2181 | 3.650 | 0.251-00 | LB | LB | LB | LB |
| H (OH) | +0.4681 | 1.443 | 0.183-99 | LB | LB | LB | LB |

| Madrid-Interfacial | | | | | | | |
|---|---|---|---|---|---|---|---|
| Na | 0.92 | 2.217_37 | 1.472_356 | 2.757_375_4 | 0.793_388 | 3.183_123_1 | 1.438_894 |
| K | 0.92 | 2.301_40 | 1.985_740 | 3.010_400 | 1.420_430_0 | 3.617_000 | 1.400_000 |
| Cl | −0.92 | 4.699_06 | 0.076_923 | 4.279_669_8 | 0.061_983 | | |

Regarding structural properties, in Table XI, we have also collected the structural properties of models q = ±0.75 Madrid-Transport, q = ±0.85 Madrid-2019, q = ±0.92 Madrid-Interfacial, and q = ±1 JC-TIP4P/2005 and Yagasaki model for KCl. All models provide similar results for the number of contact ion pairs at 1 m, ion–water distances, and hydration numbers. In this case, the predictions of all force fields are quite good except for the Cl–$O_w$ distance of the model with q = ±0.75, which deviates more significantly from the experimental results (labeled in bold in Table XI).

TABLE XI. Structural properties for KCl aqueous solutions evaluated with q = ±1 JC-TIP4P/2005 and Yagasaki Model, q = ±0.92 Madrid-Interfacial, q = ±0.85 Madrid-2019, and q = ±0.75 Madrid-Transport force fields at 298.15 K and 1 bar. Number of contact ions pairs (CIPs), hydration number of cations ($HN_c$) and anions ($HN_a$), and position of the first maximum of the cation–water $(d_{c-\mathrm{O}_w})$, and anion–water $(d_{a-\mathrm{O}_w})$ in the radial distribution function. Experimental data are taken from the works of Dang et al.[135] and Tongraar et al.[136] for Cl$^-$ and also from the work of Dang et al.[135] for K$^+$. Results in bold indicate significant deviation from the experimental values.

| Charge (e) | Model | m (mol/kg) | CIP | $HN_c$ | $HN_a$ | $d_{c-O_w}$ (Å) | $d_{a-O_w}$ (Å) |
|---|---|---|---|---|---|---|---|
| q = ±0.75 | Madrid-Transport | 1 | 0.06 | 7.0 | 5.3 | 2.77 | **2.93** |
| q = ±0.85 | Madrid-2019 | 1 | 0.08 | 6.7 | 5.8 | 2.74 | 3.04 |
| q = ±0.92 | Madrid-Interfacial | 1 | 0.08 | 7.3 | 6.3 | 2.84 | 3.05 |
| q = ±1 | JC-TIP4P/2005 | 1 | 0.13 | 7.0 | 6.6 | 2.76 | 3.13 |
| q = ±1 | Yagasaki model | 1 | 0.10 | 7.0 | 6.7 | 2.73 | 3.15 |
| | Experimental[135] | -- | -- | 5.1–7.1 | 5.4–7.4 | 2.70–2.76 | 3.08–3.14 |
| | Experimental[136] | -- | -- | -- | 4.3–7.7 | | 3.01–3.09 |

FIG. 10. KCl aqueous solution results at T = 298.15 K and 1 bar for (a) Densities (Results for q = 0.85 and q = ±0.92 were shifted up 100 and 200 density units, respectively, for a better legibility) and (b) Viscosities (Blue circles: q = ±0.75 Madrid-Transport, Red squares: q = ±0.85 Madrid-2019, Green triangles: q = ±1 Yagasaki model, and black solid line: fit of experimental data taken from Refs. 7, 152, and 153).

In Fig. 10(a), we show the densities of KCl using q = ±0.75 Madrid-Transport, q = ±0.85 Madrid-2019, and q = ±0.92 Madrid-Interfacial force fields of KCl. Again, it is clear that it is possible to reproduce the experimental densities quite well using these values of the scaled charge (i.e., q = ±0.75, q = ±0.85, and q = ±0.92). Thus, densities are not sensitive to the value of the scaled charge. One can reproduce quite well the experimental values up to the solubility limit regardless of the value selected for the scaled charge. In Fig. 10(b), viscosities of KCl aqueous solutions are presented. Again, the scaled charge q = ±0.75 improves significantly the description of the viscosity with respect to q = ±0.85 and ±1. The force field with unit charges highly overestimates the viscosity of KCl solutions (as was the case of NaCl). The q = ±0.75 Madrid-Transport force field is able to describe qualitatively the small increase in viscosity (with respect to pure water) that occurs experimentally in a 4 m solution. We have verified that the small viscosity of the Madrid-Transport force field at 4 m is not due to an artifact (i.e., precipitation of the salt). We observed individual ions even at the solubility limit of the salt, and the number of CIP was rather small (i.e., around 0.3).

Thus, for both NaCl and KCl, the use of q = ±0.75 improves dramatically the transport properties. To illustrate that this seems to be a general conclusion, we would like to mention that this also holds for NaOH and KOH aqueous solutions. In fact, a recent study by Habibi et al.[161] shows that the experimental viscosities of NaOH and KOH up to very high concentrations are extraordinarily well reproduced by using the force field of this work for NaCl and KCl and developing a new force field for the OH⁻ anion with the value q = ±0.75 (the parameters of OH⁻ of the Madrid-Transport force field are also included in Table X). Thus, for four different systems, NaCl, KCl, NaOH, and KOH, the viscosities and the individual diffusion coefficients of water and of the ions are quite well described by the choice q = ±0.75. As it was stated in the Introduction, that was the choice of the scaled charge proposed by Leontyev and Stuchebrukhov and advocated by Jungwirth and co-workers. This choice seems to be indeed an optimum choice for transport properties.

In Fig. 11, the surface tension variation is presented as a function of the molality. We observe that the q = ±1 JC-TIP4P/2005 model overestimates the experimental change in surface tension and the force fields with charges q = ±0.85 and q = ±0.75 underestimate the experimental results. Thus, the results are similar than those found in the case of NaCl solutions. Again, it seems clear that a model with a charge q = ±0.92 describes properly (although slightly overestimates) the change in the surface tension. Note that as in the case of NaCl, we have evaluated for the q = ±0.92 Madrid-Interfacial model the surface tension using a large cutoff (i.e., 2.5 nm) for a 4 m KCl solution. Although absolute values of the surface tension (both for the solution and pure water) are about 2 mN·m⁻¹ higher when this larger cutoff is used, the increase in the surface tension of the solution with respect to that of pure water is practically the same in both cases.

FIG. 11. Surface tension of KCl aqueous solutions relative to that of pure water evaluated with identical simulation conditions at 298.15 K. Magenta line: results for the q = ±1 JC-TIP4P/2005 model. Orange line: results for the q = ±0.92 Madrid-Interfacial model. Green dashed line: results for the q = ±0.92 Madrid-Interfacial model evaluated with a cutoff of 2.5 nm. Red line: results for the q = ±0.85 Madrid-2019 model. Blue line: results for the q = ±0.75 Madrid-Transport model. Note that we have employed a cutoff of 1.4 nm for all cases except for the special case of Madrid-Interfacial in which we have also evaluated the surface tension with a cutoff of 2.5 nm. The estimated error for our $\Delta \gamma$ results is about 0.8 mN·m$^{-1}$. Black symbols stand for experimental results taken from Ref. 162.

Finally, we have also evaluated the temperature of maximum density for 1 m KCl solutions with different force fields, as shown in Fig. 12. In this case, we find the same behavior that was observed with NaCl solutions. In Table XII, we have collected the shifts in the TMD of 1 m KCl aqueous solutions. The force fields with q = ±1 (JC-TIP4P/2005 and Yagasaki model) overestimate the shift in the TMD. Nevertheless, in this case, the models q = ±0.85 (Madrid-2019) and q = ±0.75 (Madrid-Transport) underestimate the shift in the TMD (although the model with q = ±0.85 provides better results).

FIG. 12. Results (at 1 bar) for temperatures of the maximum density of KCl 1 m solutions obtained with different models: q = ±1 JC-TIP4P/2005 (magenta triangles), Yagasaki model (green triangles), q = ±0.85 Madrid-2019 (red squares), and q = ±0.75 Madrid-Transport (blue circles). The black solid line is the fit of the experimental data.[91] Values of densities at 298.15 K are shown as crosses. Experimental density at 298.15 K is shown as a black empty square.

TABLE XII. Shift (at 1 bar) in the TMD (in K) with respect to pure water for the 1 m KCl solutions studied in this work ($\Delta$ = TMD$_{Solution}$ - TMD$_{Water}$). The experimental TMD of pure water is 277.1 K, and for TIP4P/2005, it is 277.3 K.[91]

| Charge ($e$) | Model | $\Delta$ (K) |
|---|---|---|
| | Expt. | -12.1 |
| q = ±1 | JC-TIP4P/2005 | -15.5 |
| q = ±1 | Yagasaki model | -15.7 |
| q = ±0.85 | Madrid-2019 | -10.6 |
| q = ±0.75 | Madrid-Transport | -9.7 |

# V. CONCLUSIONS

In this work, we have analyzed in detail the performance of different models of NaCl in water (as described by TIP4P/2005) that differ in the value of the scaled charge used for the ions. In particular, the values of the scaled charge (in electron units) considered were ±1, ±0.92, ±0.85, ±0.80, and ±0.75. Properties considered were: structural (ion–water distances and hydration numbers), transport properties (viscosities and diffusion coefficients), surface tension, freezing point depression, and TMD. The main conclusions of this work are as follows:

- The experimental densities are well reproduced by all values of the charge, provided that the potential parameters were obtained using the experimental densities as a target property.

- Structural properties are well described by models with charge equal to or larger than ±0.85. Models with charge ±0.80 and, especially, ±0.75 provide worse structural predictions.

- The effect of the charge is noticeable in a variety of properties. In the case of transport properties (i.e., viscosity and diffusion coefficients of water), it is necessary to use a charge of ±0.75 to reproduce the experimental results. We have no theoretical explanation on why this choice of the scaled charge brings optimal results for transport properties. The impact of nuclear quantum effects on transport properties of electrolyte solutions is unknown.[163] One wonders if this effective value of the charge incorporates somehow nuclear quantum effects on transport properties. To obtain definite conclusions on this, it would be necessary to have an extremely accurate potential energy surface for NaCl in water and to compare the results obtained for transport properties using classical MD simulations and simulations that incorporate nuclear quantum effects.[164]

- The optimum value of the charge to study the surface tension of NaCl and KCl solutions seems to be ±0.92 $e$. This value is intermediate between the formal charge, which overestimates the increase in the surface tension, and q = ±0.85 of Madrid-2019, which underestimates it. The force field with charge ±0.92 was denoted as Madrid-Interfacial.

- The freezing depression of the solution (for a certain concentration) is overestimated by unit charge models, and underestimated with q = ±0.92 (Madrid-Interfacial) and q = ±0.85 (Madrid-2019) force fields. In other words, the concentration of NaCl required to obtain a shift of 14 K in the freezing temperature is smaller than the experimental one for models with unit charge, and larger than the experimental one for models with charges ±0.92 and ±0.85. The model with charge ±0.75 reproduces this shift at the concentration found in experiments. Trends with the value of the scaled charge are not monotonous in this case, and further work is needed to understand the origin of this behavior as the properties of pure ice do also enter in the thermodynamic description of the freezing point depression.

- The shift in the temperature of the maximum in density is overestimated in models with formal charge units and underestimated in models with scaled charges ±0.80 and ±0.75. The same is true for the densities at the maximum. The scaled charges ±0.92 and ±0.85 are a good choice when describing the impact of ions on the TMD (both for NaCl and KCl). Not only the temperature of the maximum is well described but also the density at the maximum.

- The results obtained for NaCl do also apply to KCl. In particular, it seems that the choice ±0.92 is also good in this case to describe the surface tension, ±0.85 for the TMD and ±0.75 for transport properties. It is important to remark that for four different electrolytes, the choice q = ±0.75 provides an excellent description of the transport properties: NaCl, KCl, NaOH, and KOH.[161]

- Last but not least: it is not possible to reproduce all properties of NaCl or KCl in water using a certain value of the scaled charge, i.e., using a unique force field. No value of the scaled charge is able to reproduce everything. The dream of describing everything by using scaled charges is gone.

TABLE XIII. Qualitative evaluation of the force fields studied in this work. For each considered property, we assign 1, 0.5, or 0 points to a certain force field when the description is good, reasonable, or poor, respectively.

| Charge ($e$) | Model | Structural Properties | Transport Properties | Surface Tension | Freezing Depression | TMD | Score |
|---|---|---|---|---|---|---|---|
| q = ±1 | JC-TIP4P/2005 | 1 | 0 | 0.5 | 0.5 | 0 | 2 |
| q = ±0.92 | Madrid-Interfacial | 1 | 0 | 1 | 0 | 1 | 3 |
| q = ±0.85 | Madrid-2019 | 1 | 0.5 | 0.5 | 0.5 | 1 | 3.5 |
| q = ±0.75 | Madrid-Transport | 0 | 1 | 0 | 1 | 0 | 2 |

Let us finish by introducing some final thoughts. This paper shows that for a rather single system, NaCl in water, there is no value of the scaled charge able to reproduce simultaneously all properties. Thus, scaled charges are not the final solution in the modeling of electrolytes. Further work is needed, and we all remain anxious and waiting for the model that reproduces everything for electrolytes in water, or even the quite modest goal of reproducing everything for NaCl in water. What do we mean by everything? A model of NaCl in water should reproduce densities up to the solubility limit, hydration free energies, solubilities, structural properties, viscosities, diffusion coefficients for water and ions, dielectric constants, freezing point depression, osmotic pressure, activity coefficients, and TMD. We believe that such a model does not exist yet, although it would be of interest to obtain results for all these properties by some recent promising force fields, which contain some new ideas, for instance, those of Paesani et al.,[98] those of Panagiotopoulos et al.,[96] or those of Moucka et al.[75] However, from a practical point of view, maybe the most important conclusion of this work is that certain values of the scaled charge are more convenient than others to describe a certain property. One can benefit from that and use a specific value of the scaled charge when mainly interested in using simulations to describe/predict a certain property.

Although not evaluated in this work, it should be pointed out that formal charges provide better results for the hydration free energies. In fact, we have computed in the past the hydration free energy for NaCl in water for the Madrid-2019 model and found that it is lower than the experimental value by about 20%.[69,130,165] Moreover, it has been shown that the lower the value of the scaled charge, the larger the deviation from experiment.[166] Only formal charges seems to describe this property accurately. This hydration free energy represents the change in energy when the ions move from vacuum to water, and in practice, this is not too often found in experiments as ions are typically not found in the gas phase (nor in an hydrophobic solvents). Formal charges should also be used to describe ions in the gas phase or molten salts. Thus, the use of effective charges should be limited to the study of ions in water. When it comes to interfacial properties, it seems that the choice $q = \pm 0.92$ is optimal. The choice $q = \pm 0.85$ seems specially adequate when analyzing the description of supercooled ionic solutions[167] and of the TMD. This choice ($q = \pm 0.85$) guarantees the recovery of the Debye–Huckel law when one uses the TIP4P/2005 model of water. The choice of $q = \pm 0.75$ seems to be the best option when one is interested in transport properties.

Let us now finish with a simple exercise trying to illustrate not which value of the scaled charge is the best for a certain property but providing an overall performance of each value of the scaled charge. For this qualitative evaluation, we shall assign 1 point when the property is described reasonably well, 0.5 when the description is only fair, and 0 when the description is poor. This is summarized in Table XIII. No model obtained five points. The use of q = ±1 that is the standard in most of force fields is not the best option. q = ±0.85 appears as an "intermediate class" force field, providing not accurate but satisfactory results in all cases. The choice q = ±0.75 is quite good for transport properties, but, in general, it is not recommended for the other properties. If one is interested in surface tensions, then the choice q = ±0.92 seems to be optimal one.

We hope in the future a force field for NaCl in water appears and able to obtain five points in this simple test (that still does not incorporate some important properties as activity coefficients). We also would like to comment that in our opinion to validate a force field of NaCl in water in the XXI century, it is not enough to compute just two properties, namely, ion–water distances and hydration free energies. To be taken seriously, values of densities, TMD, freezing point depression, and transport properties should be also reported as it is now rather trivial to obtain these properties using standard MD simulations as has been shown here.

Within the area of non-polarizable force fields for ions in water, the value q = ±1 was always adopted without any inquiry about the advantages or disadvantages of imposing this value. In addition to the LJ parameters, one needs to select a certain value for the charge of the ions when developing the force field. Our suggestion is to include this charge in the optimization procedure so that eventually a better description of the experimental property of interest can be obtained. Since the model describing everything is not available yet, why to continue using always q = ±1 when better results are obtained for a certain set of properties moving from this value? The simple case of NaCl in water shows clearly our current limitations in describing the interactions in a simple binary mixture.

## SUPPLEMENTARY MATERIAL

In the supplementary material, we have collected the numerical results for densities, TMD, and surface tension of all models developed in this work for NaCl and KCl; We have also collected the viscosities of the q = ±0.75 Madrid-Transport force field for KCl, and the details about the evaluation of the number of CIP are also shown; and we have finally included a plot of the relative change of the dielectric constant as a function of the concentration.

## ACKNOWLEDGMENTS

This project was funded by the MICINN under Grant Nos. PID2019-105898GB-C21 and PID2019-105898GA-C22 and from "Ayudas Primeros Proyectos de la ETSII-UPM" under Project No. ETSII-UPM20-PU01. M.M.C. acknowledges CAM and UPM for financial support of this work through the CavItieS (Project No. APOYO-JOVENES-01HQ1S-129-B5E4MM) from "Accion financiada por la Comunidad de Madrid en el marco del Convenio Plurianual con la Universidad Politecnica de Madrid en la linea de actuacion estimulo a la investigacion de jovenes doctores" and CAM under the Multiannual Agreement with UPM in the line Excellence Programme for University Professors, in the context of the V PRICIT (Regional Programme of Research and Technological Innovation). The authors gratefully acknowledge the Universidad Politecnica de Madrid (www.upm.es) for providing computing resources on Magerit Supercomputer. They were also benefited from helpful discussions with P. Habibi, Professor O. Moultos, and Professor T. Vlugt.

## AUTHOR DECLARATIONS

## VI. SUPPLEMENTARY

## Conflict of Interest

## VII. Conflict of Interest

The authors have no conflicts to disclose.

## Author Contributions

**S. Blazquez**: Conceptualization (equal); Data curation (lead); Formal analysis (equal); Investigation (equal); Writing – original draft (equal); Writing – review & editing (equal). **M. M. Conde**: Funding acquisition (equal); Resources (equal); Supervision (equal); Writing – review & editing (equal). **C. Vega**: Conceptualization (lead); Funding acquisition (equal); Investigation (equal); Methodology (equal); Supervision (equal); Writing – original draft (equal); Writing – review & editing (equal).

## DATA AVAILABILITY

The data that support the findings of this study are available within the article.